\documentclass[letterpaper, 10 pt, conference]{ieeeconf}  

\IEEEoverridecommandlockouts                              

\usepackage{mathptmx} 
\usepackage{times} 
\usepackage{amssymb}  
\usepackage{graphicx}
\usepackage{amsmath, nccmath}
\usepackage{tabularx,ragged2e,booktabs,caption}
\usepackage{geometry}

\title{\LARGE \bf
	Designing a robust controller for a missile autopilot based on Loop shaping approach
}

\author{Li Jun Heng and  Abesh Rahman  \\
		Department of Computer Science \\
		Indiana University Bloomington \\
		Bloomington, Indiana 47401 
}

\begin{document}

	\maketitle
	\thispagestyle{empty}
	\pagestyle{empty}

\begin{abstract}
In this paper, a robust autopilot is designed for a missile autopilot, such that the system stability is guaranteed in low altitude and short-range conditions. First, using the v-gap metric, the system is linearzed around the equilibrium point. Then, the robust $H_\infty$ loop shaping controller is built for the linear model. The proposed approach does not utilize the gain scheduling method, and guarantees the system stability throughout the flight envelope. Particle Swarm Optimization (PSO) algorithm is used along with the control approach to reduce the complicated tuning process of the weight functions. The weighting functions are optimized throughout the evolutionary algorithm to maximize the stability margin. From the simulations, it is proved that the stability margins achieved guarantees the stability of interceptor throughout the whole flight envelope. 

\end{abstract}

%
\IEEEpeerreviewmaketitle

\section{Introduction}
\label{sec:introduction}
Autopilots control and design has been widely studied by the scientists and researchers [1-6]. Designing controllers for autopilots is complicated and problematic since these systems are subject to nonlinear dynamics, uncertainties, and time-varyig parameters. Therefore, in designing a controller for autopilots, not only the system stability is important, but also the system needs to operate properly in existence of parameter variations and disturbances. The disturbances are usually from the mass changes, power consumption uncertainties, and wings uncertainties, which cause the nonlinear dynamic behaviors in the system. 

Generally speaking, the gain scheduling methods are applied to design a controller for nonlinear autopilot systems, with parameters variation. In a gain scheduling method, the nonlinear model is linearized around the equilibrium points. The feedback controllers are then designed for the local linear models. Thus, utilizing the linear control theories the model, subject to small variations around the equilibrium points, can be controlled. However, if the model variations are significant, the approximate linear model no longer represents the nonlinear system's dynamics. Therefore, the classic controller fails to maintain the control performance and stability in the nonlinear system with uncertainties. One solution is to switch between different control methodologies to maintain the control objectives, however the global stability of the nonlinear system can not be proved this way. 

Gain scheduling methods based on LPV or semi-LPV approaches van guarantee system stability but they are very complicated, and they will result in conservative controllers. Thus, the best approach is to design the autopilot controller without using the gain scheduling methods, such that it can maintain control performance and stability throughout the flight envelope. For this purpose, the system equilibrium point should be defined to linearize the nonlinear model. The controller with maximum stability margin is designed then to maintain the closed loop system stability. 

$H_ \infty$ is considered as an efficient robust control approach for autopilots due to the fact that it considers both system consistency and efficiency in the control design. $H_ \infty$ loop shaping is introduced as a control method in which the uncertainties in high and low frequencies are easily modeled, and it does not deal with the complicated weighting functions and unstructured uncertainties. $H_ \infty$ loop shaping control can be applied to non-minimum phase and Multi Input Multi Output (MIMO) systems. However, the controller would be of high order and it would be hard to implement it in autopilot systems in practice. Several studies have been conducted to design lower order $H_ \infty$ loop shaping controllers for the air-crafts and autopilots. In [4], $H_ \infty$ loop shaping controller with the order equal to the weighting functions order (low order) is designed. In this method, the loop shaping weighting functions are redesigned if the stability margin was not large enough. Furthermore, the control designer needs to specify that due to which factors in the weighting functions, the stability margin is not sufficient. Then, the controller is modified to maintain stability margin and control performance. This approach is time consuming and complicated. In [5], PSO algorithm is used to optimize the weighting function parameters in the $H_ \infty$ loop shaping control procedure for a flexible beam system. A nominal model of a beam is shaped by a pre-compensator and a post-compensator to achieve a desired open loop shape. A structure-specified controller is then defined. Finally, PSO is used to search for parameters of the controller such that the cost function is minimized [5]. 

In this paper, a robust controller is designed for an autopilot system using the PSO $H_\infty$ loop shaping control procedure in [5]. Hence, the problem of high-order controller is solved and the maximum stability margin is achieved. The designed controller maintain control performance throughout the whole flight envelope. If the stability margin is not satisfactory using the selected equilibrium point, another equilibrium point is selected and the same procedure is performed. From the simulations results, the performance in tracking and system stability are satisfactory. Moreover, the studied system is non-minimum phase and this feature did not affect the system performance using this algorithm. 

The rest of the paper is organized as follows.
Section II describes system model and its parameters. 
Section III explains the proposed $H_\infty$ loop shaping control design.
Section IV provides the simulation results. 
Finally, Section V shows the conclusions and discusses future research.

\section{System Structure}
A common missile of type surface to air and air to air system is the skid-to-turn (STT) missile, in which pitch and yaw have identical response behavior. Note that the inertial cross coupling between roll, pitch, and yaw is negligible in a STT missile. The pitch yaw rotational responses behave like a spring-mass damper system. The transfer function mathematical representation of the system is expressed as (\ref{missile_model}) \cite{}.

\begin{fleqn} 
	\begin{equation}  
	\frac{\delta(c)}{\delta_c(s)}=\frac{{\omega_n}^2}{s^2+2\zeta\omega_n s+{\omega_n}^2}
	\label{missile_model}
	\end{equation}  
\end{fleqn}
where $\delta$ and $\delta_c$ are the output angle and the input angle, respectively. The natural frequency, $\omega_{n}$ is $200 \frac{rad}{sec}$, and the damping ratio, $\zeta$ is $0.7$. The equations for six degrees of freedom in the system would be as (\ref{missile_model_6degrees}) and (\ref{missile_model_6degrees_2}).

\begin{equation} 
\begingroup 
\setlength\arraycolsep{0.5pt} 
\begin{bmatrix} 
X \\
Y \\
Z
\end{bmatrix} 
= m
\begin{bmatrix} 
\dot{U}+QW-RV \\
(\dot{V}+RU) \\
\dot{W}-QU
\end{bmatrix}
\endgroup
\label{missile_model_6degrees}
\end{equation}  
\begin{equation} 
\begingroup 
\setlength\arraycolsep{0.5pt} 
\begin{bmatrix} 
L \\
M \\
N
\end{bmatrix} 
= 
\begin{bmatrix} 
I_{xx}\dot{P} \\
I_{yy}\dot{Q} \\
I_{zz}\dot{R}
\end{bmatrix}
\endgroup
\label{missile_model_6degrees_2}
\end{equation} 
where $X$, $Y$, and $Z$ are the input forces along the $x$, $y$, and $z$ axis, respectively. $P$, $Q$, $R$ are the angular speed along the $x$, $y$, $z$ axis, respectively. 

The $Y$, $Z$, $L$, $M$, and $N$ variables are calculated as (\ref{calculate}).

\begin{fleqn} 
	\begin{equation} 
	\begin{aligned}[b] 
	&Y= QS(C_{y_\beta}\beta+C_{y_{\delta_r}}\delta_r+C_{y_r}\frac{D}{2U}R)      \\
	&Z= QS(C_{z_\alpha}\alpha+C_{z_{\delta_e}}\delta_e+C_{z_q}\frac{D}{2U}Q)        \\
	&L= QSD(C_{l_{\delta_a}}\delta_a+C_{l_p}\frac{D}{2V}P)  \\
	&M= QSD(C_{m_\alpha}\alpha+C_{m_{\delta_e}}\delta_e+C_{m_q}\frac{D}{2U}Q)       \\
	&N= QSD(C_{n_\beta}\alpha+C_{n_{\delta_r}}\delta_e+C_{n_r}\frac{D}{2U}R)      
	\label{calculate}
	\end{aligned}  
	\end{equation} 
\end{fleqn}
Worth mentioning that, since the control input does not operate along the $x$ axis, the input force along the $x$ axis is chosen zero. 

After linearizing the six degrees freedom equations, the following transfer function (\ref{equation_linear} ) is attained. 

\begin{fleqn} 
	\begin{equation}  
	\frac{P_\delta(s)}{\delta a_\delta (s)}=\frac{L_{\delta a}}{sI_x - L_p}
	\label{equation_linear}
	\end{equation}  
\end{fleqn}

Moreover, the other transfer functions corresponding to the system are as (\ref{equation_1}) and (\ref{equation_2}).

\begin{fleqn} 
	\begin{equation}  
	\frac{q_\delta (s)}{\delta e_{\delta}(s)}=\frac{M_{\delta _e}s+(Z_{\delta _e}M_\alpha - Z_ \alpha M_\delta)}{s^2-(M_q+Z_\alpha /V)s+((Z_\alpha M_q-M_\alpha Z_q)/V-M_ \alpha)}
	\label{equation_1}
	\end{equation}  
\end{fleqn}
\begin{fleqn} 
	\begin{equation}  
	\frac{a_z(s)}{q_\delta (s)}=\frac{Z_\delta s^2+(M_\delta Z_q-Z_\delta M_q)s+(Z_\alpha M_\delta - Z_{\delta _e}M_\alpha)}{s^2-(M_q+Z_\alpha /V)s+((Z_\alpha M_q-M_\alpha Z_q)/V-M_\alpha)}
	\label{equation_2}
	\end{equation}  
\end{fleqn}
where the parameters in the above equations are described as (\ref{parameters}).

\begin{fleqn} 
	\begin{equation}
	\begin{aligned}[b]     
	& L_{\delta A} = QSDC_{l\delta_a}  \\
	& Z_q = \frac{SQDC_{z_q}}{m}  \\
	& L_p = QSDC_{l_p}(frac{D}{2V}) \\
	& M_\delta = \frac{SQDC_{m_{\delta_e}}}{I_y} \\
	& Z_\delta = \frac{SQC_{z_{\delta_e}}}{m}  \\
	& M_\alpha = \frac{SQDC_{m_\alpha}}{I_y}  \\
	& z_\alpha = \frac{SQC_{z_\alpha}}{m}   \\
	& M_q = \frac{SQD^2C_{m_q}}{I_yV}
	\label{parameters}
	\end{aligned}
	\end{equation}  
\end{fleqn}

\begin{figure}
	\centering
	\includegraphics[width=\columnwidth]{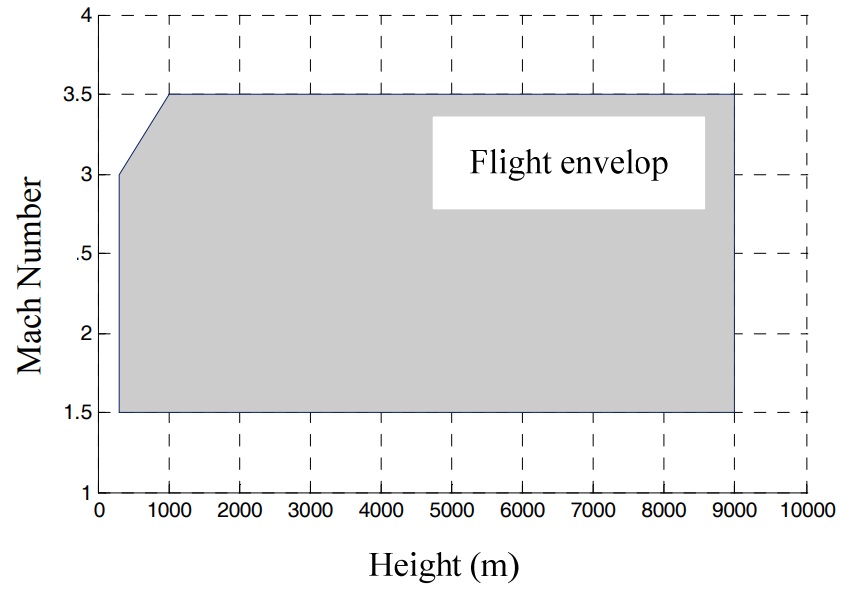}
	\caption{Flight envelope plot}
	\label{fig:envelope}
\end{figure}

The flight envelope in this study is as Fig. \ref{fig:envelope}. 

\section{Proposed $H_\infty$ Loop shaping Control Design}
To design a robust controller for the autopilot, without using the gain scheduling method, the equilibrium point of the system is needed to be defined. Then, the model is linearized around the equilibrium point. To specify the equilibrium point, v-gap metric method is used. 

\begin{figure} [h]
	\centering
	\includegraphics[width=\columnwidth]{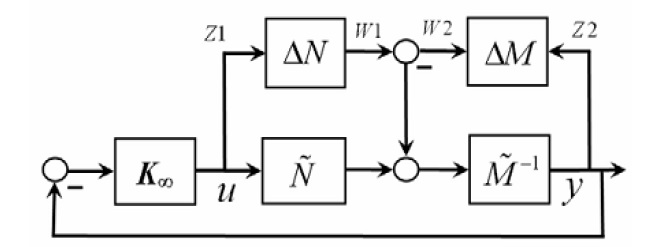}
	\caption{$H_\infty$ loop shaping control design; block diagram}
	\label{fig:loopshaping}
\end{figure}

The $H_\infty$ loop shaping controller design is based on the configuration shown in Fig. \ref{fig:loopshaping} [5]. 

The nominal model of the system is $P$, and the shaped plant with a pre-compensator $W_1$ and a post-compensator $W_2$ is defined as $P_s$. For more information see [5]. 

The open loop transfer function at the equilibrium point of the system is defined from (\ref{open_loop}).

\begin{fleqn} 
	\begin{equation}
	\begin{aligned}[b]     
	& G=\frac{k_q.G_a.\frac{q_\delta}{\delta_{e_\delta}}}{1+k_q.G_a.\frac{q_\delta}{\delta_{e_\delta}}}.\frac{a_z}{q_\delta}  \\
	& G=\frac{863878246(s-30)(s+25)}{(s+121)(s+3)(s^2+20s+7933)}
	\label{open_loop}
	\end{aligned}
	\end{equation}  
\end{fleqn}

\begin{figure} [h]
	\centering
	\includegraphics[width=\columnwidth]{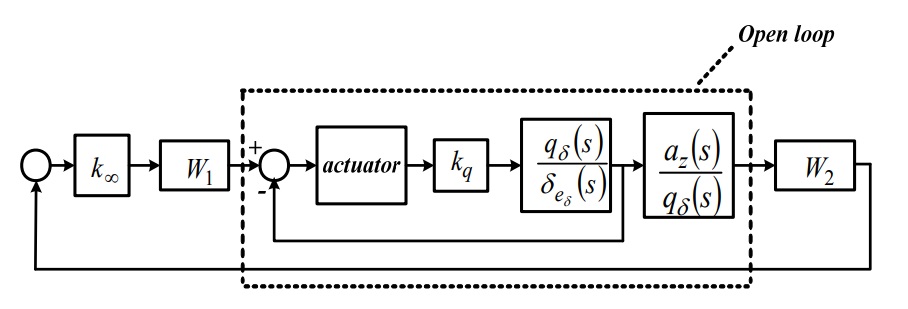}
	\caption{Block diagram of the autopilot system controlled with $H_\infty$ loop shaping control}
	\label{fig:loopshaping_malek}
\end{figure}

The open-loop transfer function is shown in the block diagram of Fig. \ref{fig:loopshaping_malek}.

\begin{figure} [h]
	\centering
	\includegraphics[width=\columnwidth]{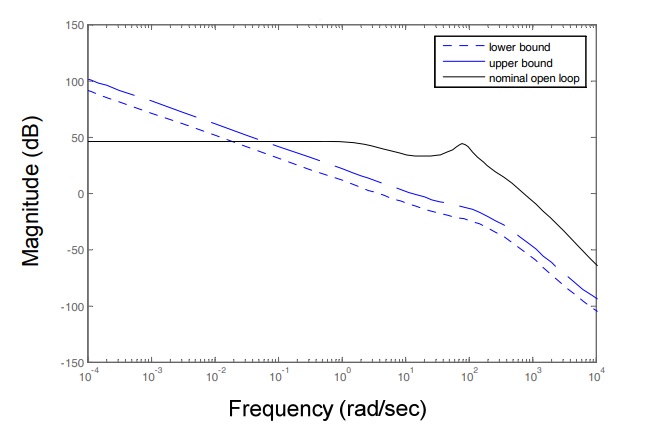}
	\caption{The frequency response of the open-loop autopilot system}
	\label{fig:ferq}
\end{figure}

The frequency response of the autopilot is shown in Fig. \ref{fig:ferq}. The controller should be designed such that it can maintain the stability and control performance of the autopilot. The open-loop gain at 300 $\frac{rad}{sec}$ should be reduced by 25 db to avoid the high-frequency dynamics and noises effects. 

The frequency limits (upper bound and lower bound) are defined as (\ref{fre_limits}).

\begin{fleqn} 
	\begin{equation}
	\begin{aligned}[b]     
	& s_{low}=\frac{3(s+40)(s+3000)}{(s+0.00001)(s+100)(s+200)(s+1000)} \\
	& s_{high}=\frac{10(s+40)(s+3000)}{(s+0.00001)(s+100)(s+200)(s+1000)}
	\label{fre_limits}
	\end{aligned}
	\end{equation}  
\end{fleqn}

So in the algorithm, the open-loop transfer function and the frequency limits (upper and lower bounds) are fed into the algorithm as the inputs. The weighting function $W_1$ is chosen as 1. The optimization runs to evaluate the weighting function $W_2$ for 100 frequency points. By mapping a minimum-phase function on these 100 points, the weighting functions are attained. 

\begin{figure} [h]
	\centering
	\includegraphics[width=\columnwidth]{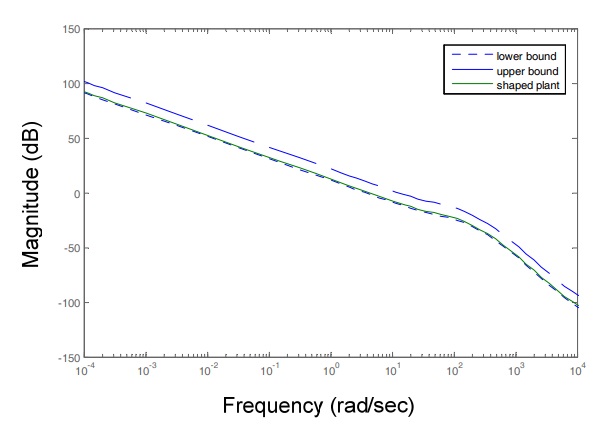}
	\caption{The frequency response of the open-loop autopilot system in the frequency limits}
	\label{fig:ferq_2}
\end{figure}

\begin{figure} [h]
	\centering
	\includegraphics[width=\columnwidth]{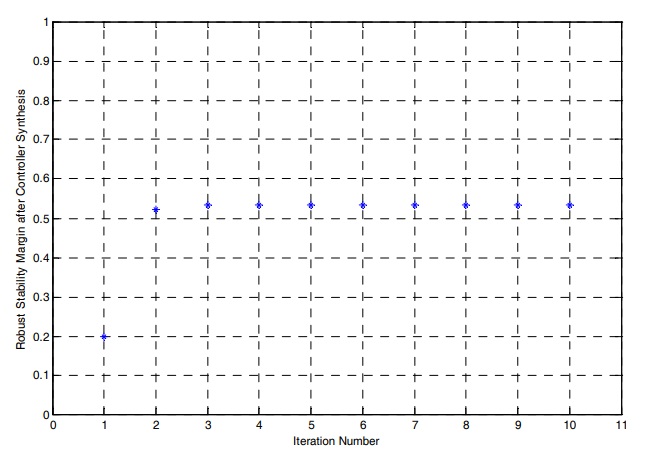}
	\caption{The stability margin of the robust controller on the autopilot}
	\label{fig:ferq_3}
\end{figure}

Fig. \ref{fig:ferq_2} presents the open-loop controlled autopilot frequency response using the frequency upper and lower bounds. The weighting functions keep the frequency response inside the frequency limits as expected. Fig. \ref{fig:ferq_3} also shows the stability margins using the $H_\infty$ loop shaping control design. Therefore, it can be understood that the designed robust controller guarantees the stability throughout the flight envelope. 

The procedure used in this paper is from [5]. The algorithm is applied to the autopilot system for the first time. The approach consists of three parts: weighting function selection, structure-specified controller, cost function selection, and the Particle Swarm Optimization algorithm. 
The weighting functions $W_1$ and $W_2$ are selected as (\ref{proposed_weights}) [5]. 

\begin{fleqn} 
	\begin{equation}
	\begin{aligned}[b]     
	& W_1 = K_1 \frac{s+\alpha_1}{s+\beta_1} \\
	& W_2 = K_2 \frac{s+ \alpha_2}{s+ \beta_2} \\
	& P_s = W_2*P*W_1
	\label{proposed_weights}
	\end{aligned}
	\end{equation}  
\end{fleqn}

The controller $K(s)$ is a fixed-structure controller, with $n$ inputs and $m$ outputs, as (\ref{controller_fixed}). 

\begin{fleqn} 
	\begin{equation}
	K(s) = \frac{a_ms^m + a_{m-1}s^{m-1}+ \cdots + a_0}{s_n + b_{n-1}s^{n-1} + \cdots + b_0}
	\label{controller_fixed}
	\end{equation}  
\end{fleqn}

The order of the controller chosen is arbitrary, based on the system details and specifications. Once the controller is defined, the parameters of the controller and the weighting functions are optimized using the Particle Swarm Optimization (PSO) algorithm [5].

A more exact method is to use singular identification and estimation methods to identify the system parameters (since the autopilot systems usually include singularities and singular eigenvalues), and then use the proposed methodology [7].

\section{Simulation Results}
\begin{figure} [h]
	\centering
	\includegraphics[width=\columnwidth]{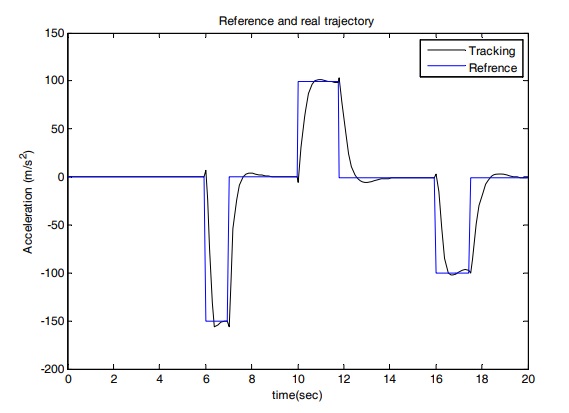}
	\caption{input signal applied to the autopilot system}
	\label{fig:input}
\end{figure}

\begin{figure} [h]
	\centering
	\includegraphics[width=\columnwidth]{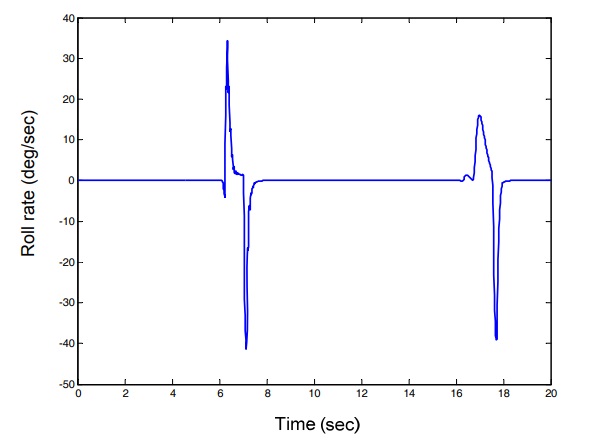}
	\caption{Angular frequency in the autopilot system}
	\label{fig:input2}
\end{figure}

The designed $H_\infty$ loop shaping controller is implemented on a autopilot with six degrees of freedom. The controlled system maintain the system performance and stability. The reference signal is applied to the system as a step function as Fig. \ref{fig:input}. The system's output and reference plots are shown in Fig. \ref{fig:input}. As expected, the tracking is satisfactory. Fig. \ref{fig:input2} shows the angular speed that should converge to zero. 

\begin{figure} [h]
	\centering
	\includegraphics[width=\columnwidth]{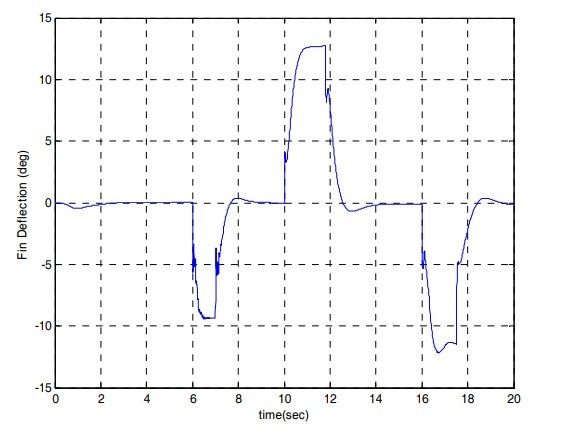}
	\caption{Control input signal}
	\label{fig:input3}
\end{figure}

\begin{figure} [h]
	\centering
	\includegraphics[width=\columnwidth]{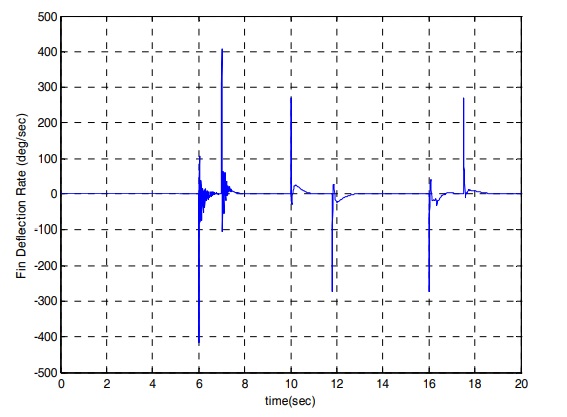}
	\caption{Derivative of the control input signal}
	\label{fig:input4}
\end{figure}

Figs. \ref{fig:input3} and \ref{fig:input4} show the control input signal and the derivative of the control signal, respectively.

In Table \ref{tab:results}, the autopilot performance using the robust controller are represented. Based on the values, the performance and the stability of the system are satisfactory. 

\begin{table}
	\begin{center}
		\caption{Traffic network Parameters}
		\begin{tabular}{| c | m{1cm}|m{1cm}|m{1cm}|m{1cm}|}
			\hline
			& {position 1} & {position 2} & {position 3} & {position 4}  \\  \hline
			overshoot time & 0.36 & 0.64 & 0.69 & 0.75  \\ \hline
			Overshoot & 0.04 & 0.03 & 0.02 & 0.06  \\ \hline
			steady state error & 0.01 & 0.01 & 0.01 & 0.01  \\ \hline
			maximum rate of angle change & 27.5 & 27.5 & 27.5 & 27.5  \\ \hline
		\end{tabular}
		\label{tab:results}
	\end{center}
\end{table}

\section{Conclusion and Future Work}
In this paper,  robust controller is designed and implemented in an autopilot system. The simulation results on a autopilot system, with six degrees of freedom, showed that the controlled system guarantees the stability throughout the whole flight envelope and performs satisfcatorily in tracking the reference signal. In the proposed approach, the equilibrium point is specified, then the system is linearized around the equilibrium point, and them the $H_\infty$ loop shaping controller is designed by optimizing the weighting functions through PSO algorithm. The complication and hard problem of defining the weighting functions for the autopilot system are solved through the approach. The controller designed is also reasonable in practice since it is not of very high order. For future work, combining the used approach with the adaptive control design is recommended.

\end{document}